\documentstyle[12pt,emulateapj,psfig]{article}

\newcommand{\gtilde}
 {~ \raisebox{-1ex}{$\stackrel{\textstyle >}{\sim}$} ~}

\def\ltsima{$\; \buildrel < \over \sim \;$}
\def\ltsim{\lower.5ex\hbox{\ltsima}}
\def\gtsima{$\; \buildrel > \over \sim \;$}
\def\gtsim{\lower.5ex\hbox{\gtsima}}

\begin{document}
\submitted{To Appear in ApJ Letters}

\title{Evolution of Dust Extinction and Supernova Cosmology}

\author{Tomonori Totani$^1$ and Chiaki Kobayashi$^2$}
\affil{ $^1$National Astronomical Observatory, Mitaka, Tokyo 181-8588,
Japan (E-mail: totani@th.nao.ac.jp) \\
$^2$Department of Astronomy, School of Science,
University of Tokyo, Bunkyo-ku, Tokyo 113-0033, Japan;
 (E-mail: chiaki@astron.s.u-tokyo.ac.jp)}

\footnotesize

\begin{abstract}
We have made a quantitative calculation for the systematic evolution
of average extinction by interstellar dust in host galaxies of high-redshift
Type Ia supernovae, by using a realistic model of photometric and
chemical evolution of galaxies and supernova rate histories 
in various galaxy types.
We find that average $B$ band extinction
$\langle A_B \rangle$ at $z \sim$ 0.5 is typically 0.1--0.2 mag larger than
present, under a natural assumption that dust optical depth
is proportional to gas column density and gas metallicity.
This systematic evolution causes average reddening with
$E(B-V) \sim$ 0.025--0.05 mag with the standard extinction curve,
and this is comparable with the observational uncertainty 
of the reddening of high-redshift supernovae. Therefore, our result does not
contradict the observations showing
no significant reddening in high-$z$ supernovae.
However, the difference in apparent magnitude 
between an open universe and a $\Lambda$-dominated
flat universe is only $\sim$ 0.2 mag at $z\sim$0.5, and hence this systematic
evolution of extinction should be taken into account in a reliable
measurement of cosmological parameters. Considering this 
uncertainty, we show that it is difficult to discriminate between an open
and $\Lambda$-dominated flat cosmologies from the current data.
\end{abstract}


\section{Introduction}
Type Ia supernovae (SNe Ia) have been known as a representative
standard candle in the universe, and used in measurements of
cosmological parameters such as the Hubble constant ($H_0$), 
density parameter ($\Omega_M$), and the cosmological constant
($\Omega_\Lambda$). Recently two independent groups have obtained
the same result that a $\Lambda$-dominated flat universe is strongly
favored, by using a few tens of SNe Ia at redshift 
$z \sim$ 0.5 (Riess et al.
1998 [R98]; Perlmutter et al. 1999 [P99]). Extinction by dust could affect
these analyses, and both groups have made a considerable effort
to assess the systematic uncertainty due to extinction. Both groups
reported that there is no significant color difference 
between the high-$z$ and local samples. In the sample of P99, 
average reddening $\langle E(B-V) \rangle$ is 0.033$\pm$0.014 mag
for the local sample and 0.035$\pm$0.022 for the high-$z$ sample (P99).
The mean $B-V$ color of the high-$z$ R98 sample
is $-0.13\pm$0.05 or $-0.07\pm$0.05 depending on two analysis methods,
while expectation of unreddened color is $-0.10$ to $-0.05$. 
However, there is a statistical uncertainty of $\gtilde 0.03$ mag
in the color difference, and 
in addition, there is a systematic uncertainty in the
K-correction of about 0.03 mag (P99, and probably similar number also
for R98). Therefore, there is observational uncertainty of at least
$\sim$ 0.05 mag in reddening evolution for both
groups. Furthermore, these error estimates have been achieved by
statistical averaging of many supernovae; typical color uncertainty
for each supernova is much larger ($\sim 0.1$--0.2 mag, see Fig.
6 of P99). Therefore, it is difficult to clearly check 
a systematic evolution of average $B$-band 
extinction $\Delta A_B \sim R_B \Delta E(B-V) \sim 0.2$ mag
with a reddening of $\Delta E(B-V) \sim 0.05$ by the current
observational data, where $R_B$ is the total-to-selective extinction ratio
for the $B$ band,
and $R_B \sim 4$ for the standard extinction curve. On the other hand,
the difference of apparent magnitude between 
an open universe and a $\Lambda$-dominated flat universe
is only $\sim$ 0.2 mag at $z \sim 0.5$, and hence it is still possible
that unchecked systematic evolution of extinction has affected
the measurements of cosmological parameters. 

Aguirre (1999a, b) considered the effect of intergalactic dust
which was ejected from galaxies. Such dust could have a significantly
grey extinction and Aguirre has shown that this kind of dust may affect 
measurements of cosmological parameters with smaller reddening. 
This is an interesting possibility, but existence of such 
intergalactic dust has not yet been confirmed. Here we consider the
dust existing in host galaxies with the standard extinction curve.
Although it is uncertain whether supernovae evolve to high redshifts, 
their host galaxies should undoubtedly evolve as shown by
various observations of galaxies at high redshifts. Both the
gas column density and gas metallicity, which are important physical
quantities for dust opacity, change with time by various star formation
histories depending on morphological types of galaxies.
Therefore, average dust extinction
in host galaxies should evolve systematically, and the aim of this paper is to
make a quantitative estimate for this evolution, by using
a realistic model of photometric and chemical evolution of galaxies
and supernova rate histories in various types of galaxies. We find that
typical evolution in average $A_B$ is $\sim$ 0.1--0.2 mag from $z=0$
to $\sim$0.5, 
which is significant for measurements of cosmological parameters
but may have escaped from the reddening check. Therefore, this effect 
should not be ignored in measurements of cosmological parameters
by high-$z$ SNe Ia. 

\section{Evolution of Average Extinction in Host Galaxies}
In this letter we consider only the average extinction of a supernova in a
host galaxy, and do not consider the variation within a galaxy
depending on the supernova location in it.
Although the variation within a host galaxy can be washed out
by statistical averaging of many supernovae, evolution of galaxies
will cause systematic evolution of average extinction which cannot
be removed by statistical averaging. It is physically natural to 
assume that the dust-to-metal ratio is constant and hence 
the dust opacity is proportional to gas column density and 
gas metallicity of a host galaxy. In fact, it is well known that
the extinction in our Galaxy is well correlated to 
the HI gas column density (e.g., Burstein \& Heiles 1982; Pei 1992). 
It is also known that the dust opacity is correlated
to the metallicity among the Galaxy and the Large and Small
Magellanic Clouds, when gas column density is fixed (e.g., Pei 1992). 
Hence in the following we assume that the
dust opacity is proportional to gas column density and
gas metallicity, which evolve according to the star formation
history in a galaxy. 

The star formation history can be inferred from the present-day
properties of observed galaxies, by using the well-known technique of stellar
population synthesis. We can estimate the
time evolution of gas fraction and metallicity in a galaxy by using
photometric and chemical evolution models
for various galaxy types. Since the observed extinction of high-$z$
SNe Ia is an average over various types of galaxies,
we also need the evolution of SN Ia rate in various galaxy types.
In the next section, we describe the model of galaxy evolution and
SN Ia rate evolution used in this letter, which is constructed to
reproduce various properties of the present-day galaxies.

\subsection{Evolution of galaxies and Type Ia supernova rate}
We use photometric and chemical evolution models for five morphological types
of E/S0, S0a-Sa, Sab-Sb, Sbc-Sc, and Scd-Sd. The basic framework of the model
is the same as that of elliptical galaxies of Arimoto \& Yoshii
(1987) and that of spiral galaxies of Arimoto, Yoshii, \& Takahara (1992),
but model parameters are updated to match the latest observations
(Kobayashi et al. 1999), by
using an updated stellar population database of Kodama \& Arimoto (1997)
and nucleosynthesis yields of supernovae of Tsujimoto et al. (1995).
The model parameters for spiral galaxies are determined to reproduce 
the present-day gas fractions and $B-V$ colors in various galaxy types 
at 15 Gyr after the formation.
The model of elliptical galaxies is the so-called galactic wind model,
in which star formation stops at about 1 Gyr after the formation
by a supernova-driven galactic wind (\cite{lar74}; \cite{ari87}). 
We assume that 
gas fraction in a elliptical galaxy decreases exponentially
after the galactic wind time ($\sim$ 1 Gyr), with a time scale same as
the galactic wind time. These models give the evolution of gas fraction
and metallicity in each galaxy type depending on the star formation
history.

SN Ia rate history in each type of galaxies is calculated with 
the metallicity-dependent SN Ia model introduced by Kobayashi et al. (1998).
In their SN Ia progenitor model, an accreting white dwarf (WD) 
blows a strong wind to reach the Chandrasekhar mass limit.  
If the iron abundance of progenitors is as low as [Fe/H]$\ltsim -1$, 
the wind is too weak for SNe Ia to occur. 
Their SN Ia scenario has two progenitor systems: one is a red-giant
(RG) companion with the initial mass of 
$M_{\rm RG,0} \sim 1 M_\odot$
and an orbital period of tens to hundreds days (\cite{hkn96}, 1999).
The other is a near main-sequence (MS) companion with an initial mass of
$M_{\rm MS,0} \sim 2$--$3 M_\odot$
and a period of several tenths of a day to several days 
(\cite{li97}; \cite{hknu99}).
The occurrence of SNe Ia is determined from two factors: lifetime of
companions (i.e., mass of companions) and iron abundance of progenitors.
(See Kobayashi et al. 1998, 1999 for detail.)
This model successfully reproduces the observed chemical evolution in
the solar neighborhood such as the evolution of oxygen to iron ratio
and the abundance distribution function of disk stars (\cite{kob98}),
the present SN II and Ia rates in spirals and ellipticals, and
cosmic SN Ia rate at $z \sim 0.5$ (\cite{kob99}).

\subsection{Average extinction evolution towards high redshifts}
We have modeled the evolution of gas fraction ($f_{\rm g}$), 
metallicity ($Z$), and SN Ia rate per unit baryon mass of a galaxy
(${\cal R}_{Ia}$) in various types of galaxies, from which we calculate
the evolution of average extinction in the universe. 
We assume that these quantities do not depend on the mass of galaxies.
The basic assumption is that the dust opacity, and hence average
$A_B$ in a galaxy is proportional to gas column density and
gas metallicity. The average extinction
at redshift $z$ in a $i$-th type galaxy with the present-day $B$ luminosity
$L_B$ is given by $A_{B,i}(z, L_B) = \kappa f_{{\rm g},i}(t_z)
Z_i(t_z) [r_{{\rm e}, i}(L_B)]^{-2}  (M_{\rm b}/L_B)_i L_B$, where $t_z$ is 
the time from formation of galaxies,
$r_{\rm e}$ the effective radius of galaxies, and $(M_{\rm b}/L_B)$ is
the baryon-mass to light ratio which is determined by the evolution model.
We assume a single formation epoch $z_F$ for all galaxy types for 
simplicity.\footnote{In reality, there should be some dispersion in galaxy
ages. However, the systematic evolution of dust extinction 
is owing to the fact that all galaxies should become younger on average
towards high redshifts, and present-day 
age dispersion cannot remove this systematic 
effect.} The proportional constant $\kappa$ will be determined later.
We do not consider the size evolution of galaxies, and determine
$r_{\rm e}(L_B)$ from
empirical relations observed in local galaxies (Bender et al. 1992 for 
ellipticals, and Mao \& Mo 1998 for disk galaxies). 
It should be
noted that the extinction depends on the absolute luminosity of
galaxies. From the empirical $L_B$-$r_{\rm e}$ relation, the surface
brightness becomes brighter with increasing luminosity of disk galaxies,
and hence the massive galaxies should be more dusty than smaller ones.
This trend is consistent with observations (van den Bergh and Pierce
1990; Wang 1991).
Then the average extinction of SNe Ia over all galaxy types
at a given redshift is
\begin{eqnarray}
\langle A_B(z) \rangle = 
\frac{ \sum_i \int dL_B 
A_{B, i}(z, L_B)
{\cal R}_{{\rm Ia}, i}(t_z) (M_{\rm b}/L_B)_i L_B \phi_i(L_B)}{
\sum_i \int dL_B {\cal R}_{{\rm Ia}, i}(t_z) (M_{\rm b}/L_B)_i L_B 
\phi_i(L_B)} 
\nonumber
\end{eqnarray}
where $\phi_i$ is the type-dependent galaxy luminosity function at
$z=0$, for which we adopted the Schechter parameters
derived by Efstathiou, Ellis, \& Peterson (1988)
using the catalog of the Center for Astrophysics (CfA) Redshift Survey
(Huchra et al. 1983).

We have to determine the overall normalization of extinction, $\kappa$, 
for which we use the average $V$ extinction of the Milky Way
(Sbc type, $L_B = 1.4 \times 10^{10}L_{B\odot}$) at $z=0$: $\langle
A_V \rangle_{\rm MW}$. This is an average of extinction of SNe Ia
occurring in our Galaxy seen by an extragalactic observer, and hence
it is different, in a strict sense, from the
average of the Galactic extinction which is extinction
of extragalactic objects observed by us. However,
if the location of the Sun is typical in the Milky
Way, we may infer this quantity by the average of the Galactic extinction.
The average Galactic extinction
of the 42 SNe Ia observed by P99 is $\sim$ 0.1 mag in $A_R$
or $A_I$ (see Table 1 of P99).
This suggests $\langle A_V \rangle_{\rm MW} \sim 0.1$--0.2 with the
standard Galactic extinction law (e.g., Pei 1992). 
The average reddening of the
Galaxy then becomes $\langle E(B-V) \rangle_{\rm MW} \sim$ 0.03-0.06 mag,
which is a typical reddening at the Galactic latitude of
$\sim$40--50$^\circ$ in the Galactic extinction map (Burstein \& Heiles 1982;
Schlegel, Finkbeiner, \& Davis 1998). This estimate is consistent
with a model of dust distribution in our Galaxy, which
suggests that the average extinction of SNe Ia 
in the Galaxy seen by an extragalactic observer
is typically $\langle A_V \rangle_{\rm MW} \sim$ 0.1--0.2 mag
(Hatano, Branch, \& Deaton 1998).\footnote{
Hatano et al. suggested that
most supernovae are only mildly obscured but
there is a long tail to stronger extinctions in the extinction
distribution. In the actual observations, 
such a tail will be cut out due to a magnitude limit of a survey. 
Hence, we have used here the mean extinction
of the ``extinction-limited subset'' in the Table 1 of Hatano et al., in which
strongly obscured supernovae with $A_B>0.6$ are removed.} 
Therefore we use $\langle A_V
\rangle_{\rm MW} \sim$ 0.1--0.2 mag as a plausible range of the
average extinction of our Galaxy.

Figure 1 shows the evolution of $B$ extinction for each galaxy type
as well as the average over all galaxy types, normalized by 
$\langle A_V \rangle_{\rm MW}$,
i.e., $\langle A_B(z) \rangle/\langle A_V \rangle_{\rm MW}$. 
Here we used a cosmological model with
$(h, \Omega_M, \Omega_\Lambda) = (0.5, 0.2, 0)$, and set $z_F = 4.5$
so that the age of galaxies is 15 Gyr which was assumed in
the evolution model. The thick solid line is
the average over all galaxy types, and the thin lines are for individual
galaxy types as indicated.
Since we have used a galactic wind model for elliptical galaxies,
they do not have interstellar gas and hence there is
no extinction in elliptical galaxies at $z<1$. The evolution of extinction is
caused by spiral galaxies, but the behavior of evolution is considerably
dependent on galaxy types. Early-type spiral galaxies become
more dusty towards $z \sim 1$, but an opposite trend is seen for late types. 
These behaviors can be understood as a competition of
the two effects: gas fraction evolution and metallicity evolution.
The gas fraction in early spiral galaxies is much smaller
than late types at present, but rapidly increases towards high
redshifts. This increase is responsible for increase of gas column
density and hence the dust opacity. On the other hand, the gas fraction
does not increase so much in late type galaxies, and decrease of metallicity 
towards high redshifts is responsible for the decrease
of dust opacity. In redshifts
more than 1, the extinction decreases towards higher redshifts 
in all spiral galaxies because the metallicity evolution becomes
dominant. 

The average over all types is weighted by the SNe Ia rate
in each type. Because the star formation rate increases more rapidly
to high redshifts in early-type spiral galaxies than late types,
the average extinction is more weighted to early types at higher 
redshifts. Hence $\langle A_B \rangle / \langle A_V \rangle_{\rm MW}$ 
increases to high redshifts by $\sim 1$ from $z=0$ to 0.5.
This result suggests that, with $\langle A_V \rangle_{\rm MW} \sim$
0.1--0.2, the average extinction $\langle A_B \rangle$ 
of SNe Ia at $z \sim 0.5$ is larger than the local sample by about
0.1--0.2 mag. This systematic evolution of average extinction
is comparable with the difference between an open and
a $\Lambda$-dominated universe in the Hubble diagram, 
and hence this effect significantly
affects measurements of cosmological parameters. In the next section
we apply the above model in the estimate of cosmological parameters
by using the sample of P99.

\section{Effect on the Cosmological Parameters}
Figure 2 shows the Hubble diagram for SNe Ia of the
primary fit C of P99, which plots restframe 
$B$ magnitude residuals from a $\Lambda$-dominated flat 
cosmology [$(h, \Omega_M, \Omega_\Lambda) = (0.65, 0.2, 0.8)$] without
dust effect (thin solid line). 
Thin long- and short-dashed lines are the predictions
of the dust-free case with an open universe (0.5, 0.2, 0.0) and the
Einstein-de Sitter (EdS) universe (0.5, 1.0, 0.0), respectively. 
As reported by P99, the 
$\Lambda$-dominated flat universe gives the best-fit to the data.
Next, the thick lines show the predictions when the model of 
extinction evolution is taken into account, where
the cosmological parameters are the same with the dust-free curves
of the same line-markings. Here we adopt 
$\langle A_V \rangle_{\rm MW}$ = 0.2 mag.
In the open and $\Lambda$-dominated models, the galaxy formation epoch
is set to $z_F = 4.5$ and 5.0 so that the age of galaxies becomes
15 Gyr. In the EdS model, the age of the universe is shorter than 15 Gyr,
and hence we set $z_F = 5$ which gives an age of galaxies as
$\sim$ 12 Gyr. Although this age is a little shorter than that
assumed in the evolution model, the evolutionary effect
during 12--15 Gyr is small and hence this inconsistency is not serious.
As expected, the model curves with the dust effect are typically 
0.1--0.2 mag fainter than those without dust. As a result,
the open universe becomes the most favored cosmology among the three
when the extinction evolution is taken into account. 

We avoid more detailed statistical analysis to 
derive any decisive conclusion about the cosmological parameters,
because the result would be highly dependent on the
extinction evolution model. However, the evolution model presented
here is quite a natural and standard one without any exotic assumption.
Hence, our conclusion is that 
the systematic evolution of average extinction in host galaxies
should more carefully be taken into account when one uses 
SNe Ia to constrain the cosmological parameters.

\section{Discussion}
P99 estimated the systematic uncertainty of extinction to be
less than 0.025 mag ($1\sigma$) in $A_B$, 
based on an analysis after removing nine reddest supernovae. 
Aguirre (1999b) argued that this limit does not apply if 
the dispersion in brightness and/or colors of high-$z$ supernovae
is dominated by factors other than extinction. 
As noted before, the observational 
uncertainty in $E(B-V)$ for each supernova is typically $\sim 0.1$--0.2
mag, which is larger than the systematic reddening
of $E(B-V) \sim$ 0.025--0.05 considered in this letter.
Therefore, it is doubtful that P99 analysis successfully 
removed high-$z$ supernovae reddened by the systematic 
extinction evolution. Rather, 
the supernovae removed by P99 might be reddened by strong reddening
depending on locations in host galaxies, or by some other reasons
as argued by Aguirre (1999b). Hence,
we consider that the systematic evolution of dust considered here
has not yet been checked by the observations.

In the analysis of R98, the reddening correction is systematically
included in the process of light-curve-shape fitting, and
one may consider that our result does not apply to the analysis of R98.
However, as noted in Introduction, one must correct a reddening
with $E(B-V) \sim$ 0.05 to discriminate between an open and $\Lambda$
cosmologies, and this is comparable with the observational error
of color estimates.
In principle, it is difficult to correct the reddening effect
when the reddening is as small as the observational error in colors,
because the reddening correction is performed based on the observed colors.
The reddening correction by R98 will be effective for 
supernovae strongly reddened beyond the color uncertainty,
but it is not clear to us whether this correction has successfully corrected 
the systematic reddening evolution discussed in this letter.

The two groups (R98, P99) argued that the observed dispersion of apparent 
magnitudes showing no significant evolution to high redshifts
gives further support that their results are not affected by
extinction. Their argument on the dispersion
test is true if the observed dispersion is dominated by the
dispersion of extinction, and R98 claimed that the observed dispersion 
is smaller than that expected from a dust
distribution model of Hatano et al. (1998) if mean extinction is
$\sim$ 0.2 mag. However, there is large uncertainty
in dust distribution models, and the selection effect might have reduced the
apparent dispersion (see the footnote 2).
We should also be careful about the uncertainty in the observed
`intrinsic' dispersion, which was derived by subtracting the measurement
error from the actually observed dispersion (P99). Hence, this test does 
not favor our scenario but cannot strictly reject it.

We suggest that the best way to constrain the cosmological parameters
by high-$z$ SNe Ia would be an analysis using only 
SNe Ia in elliptical galaxies, in which the dust evolution effect
is much smaller than spiral galaxies at $z<1$. 
In fact, P99 tried to analyze their supernovae
with known host-galaxy types, and found no significant change
in the best-fit cosmology. However, the host-galaxy classification
is only based on spectra of host galaxies without high-resolution
images. The uncertainty of a fit with a specified host-galaxy 
type is still large due to the limited number of supernovae, 
and P99 concluded that this test will need to await the host-galaxy
classification of the full set of high-$z$ supernovae and
a larger low-$z$ supernova sample. It is expected from our 
calculation that the fitting result of such an analysis in the future
will be dependent on host-galaxy types, giving important information
for chemical evolution of galaxies. High-$z$ supernovae beyond
$z \sim 1$ are also desirable to study galaxy evolution
as well as cosmological parameters.

\begin{figure}
  \begin{center}
    \leavevmode\psfig{file=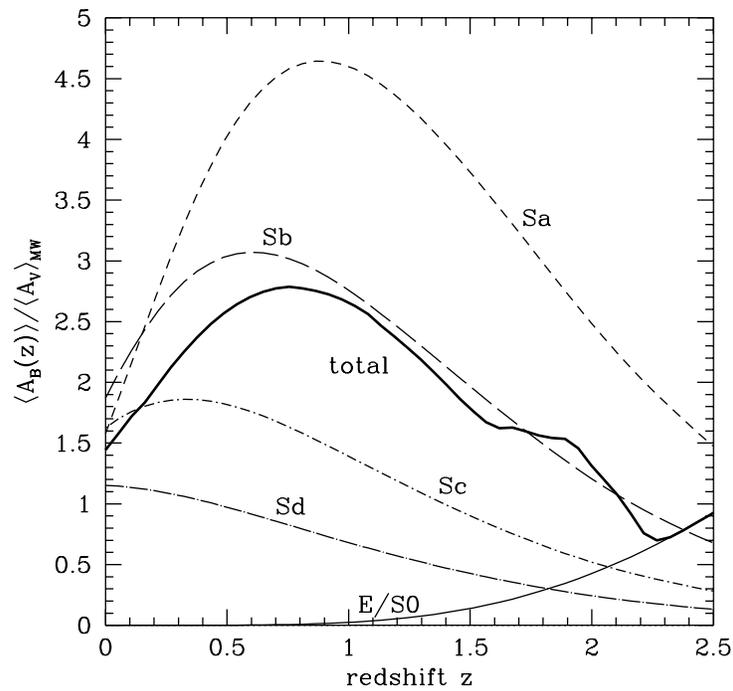,width=10cm}
  \end{center}
\caption{Average $B$-band extinction of type Ia supernovae
as a function of redshift. Extinction $A_B$ is normalized by
the average $V$ extinction of our present-day
Galaxy, $\langle A_V \rangle_{\rm MW}$ 
(see text). The thick solid line is the average over the five morphological
types of galaxies, which is weighted by SN Ia rate in them.
Five thin lines are extinction evolution in individual galaxy types, as
indicated. An open universe with $(h, \Omega_M, \Omega_\Lambda) =
(0.5, 0.2, 0.0)$ is assumed, and the formation epoch of galaxies
is set to $z_F = 4.5$.}
\end{figure}

\begin{figure}
  \begin{center}
    \leavevmode\psfig{file=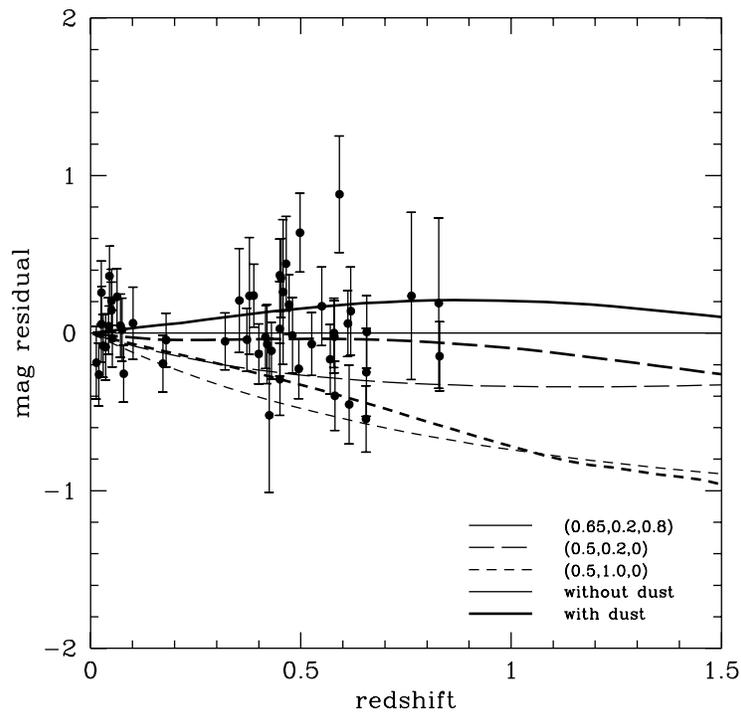,width=10cm}
  \end{center}
\caption{The Hubble diagram of type Ia supernovae. The data are those
used in the primary fit C of Perlmutter et al. (1999). 
Restframe $B$ Magnitude residuals are from a $\Lambda$-dominated universe 
with $(h, \Omega_M, \Omega_\Lambda) = (0.65, 0.2, 0.8)$ without dust
effect (thin solid line). Thin long- and short-dashed lines 
are predictions
of an open universe (0.5, 0.2, 0.0) and Einstein-de Sitter universe
(0.5, 1.0, 0.0) without dust. The three thick lines are predictions
for the case with dust evolution, where the cosmological parameters
are the same with the dust-free curves of the same line-markings.
}
\end{figure}

\end{document}